\documentclass[final,5p,times,twocolumn]{elsarticle}
\usepackage{amssymb}
\usepackage{lineno}
\usepackage[utf8]{inputenc}
\usepackage[T1]{fontenc}
\usepackage{epsfig}
\usepackage{amsmath}
\usepackage{amsfonts}
\usepackage{color}
\usepackage{graphicx}
\usepackage{url,hyperref}
\usepackage{latexsym}
\usepackage{longtable}
\usepackage{float}
\usepackage{epstopdf}
\usepackage{url}
\usepackage{caption}

\usepackage{mathrsfs}
\usepackage[titletoc]{appendix}
\usepackage{natbib}
\usepackage{textcase}
\usepackage{bm}

 \newcommand{\bq}{\begin{equation}}
 \newcommand{\eq}{\end{equation}}
 \newcommand{\bqn}{\begin{eqnarray}}
 \newcommand{\eqn}{\end{eqnarray}}
 \newcommand{\bwt}{\begin{widetext}}
 \newcommand{\ewt}{\end{widetext}}
 \newcommand{\nb}{\nonumber}
 \newcommand{\lb}{\label}

\newcommand{\Lag}{\textbf{L}}
\newcommand{\lag}{\mathcal{L}}
\newcommand{\Lie}{\pounds}
\newcommand{\ud}{\mathrm{d}}

\newcommand{\F}{\mathcal{F}}
\newcommand{\A}{\mathcal{A}}
\newcommand{\Q}{\textbf{Q}}
\newcommand{\J}{\textbf{J}}
\newcommand{\B}{\textbf{B}}

\newcommand{\AEt}{\text{\AE}} 
\newcommand{\aet}{\text{\ae}}
\newcommand{\ux}{(u\cdot \xi)}

\begin{document}
\begin{frontmatter}
\title{Smarr Integral Formula of D-dimensional Stationary Spacetimes in
Einstein-\AE ther-Maxwell Theroy}

\author{Fei-hung Ho $^{a,c}$}
\author{Shao-Jun Zhang $^{a}$}
\author{Haishan Liu $^{a}$}
\author{Anzhong Wang\corref{cor1}$^{a,b}$}
\ead{Anzhong\_Wang@baylor.edu}
\cortext[cor1]{Corresponding author}
\address{
$^{a}$ {Institute for Advanced Physics \& Mathematics, Zhejiang University of Technology, Hangzhou, 310023, China}\\
$^{b}$ {GCAP-CASPER, Physics Department, Baylor University, Waco, Texas 76798-7316, USA}\\
$^{c}$ {Department of Physics, Jimei University, Xiamen, 361021, China}
}

\begin{abstract}

Using the Wald formalism, we investigate the thermodynamics of charged black holes in D-dimensional stationary
spacetimes with or without rotations in Einstein-\ae ther-Maxwell theory. In particular, assuming the existence of a 
scaling symmetry of the action, we obtain the Smarr integral formula, which can be applied to both Killing and 
universal horizons. When restricted to 4-dimensional spherically symmetric spacetimes, previous results obtained 
by a different method are re-derived.

\end{abstract}

\end{frontmatter}
%


\section{Introduction}
\renewcommand{\theequation}{1.\arabic{equation}} \setcounter{equation}{0}

Jacobson and Mattingly introduced a unit time-like 4-vector field $u^a$, the so-called  ``\ae ther field'' coupled to Einstein's general relativity (GR), which is often referred to as the Einstein-\ae ther theory \cite{JM01}.
The existence of the \ae ther field defines a preferred frame, whereby the Lorentz invariance is locally broken and leads to some novel
effects: superluminal gravitational  modes with different speeds exist \cite{JM04}.
Due to the existence of the superluminal gravitational modes, the causality of the theory will be quite different from that of GR.
In particular, since the speeds of these particles in principle can be arbitrarily large, so the corresponding light-cones can be completely flat, and the causality of the theory is more like that of Newtonian theory \cite{GLLSW,Wang17}.
Then, some natural questions rise: does black holes still exist? how to define a black hole now?

To answer these questions, we need first to understand the corresponding causality of the theory.
In Newtonian theory, an absolute time $t$ is first introduced, and the causality is guaranteed by assuming that all particles move in the
increasing direction of $t$.
With this in mind, Blas and Sibiryakov first realized that the time-like \ae ther field plays the role of the absolute time of Newton \cite{BS11}.
In particular, the \ae ther field naturally defines a global timelike foliation of a given spacetime, and all the particles are assumed to move along the increasing direction of the foliation.
Once the causality of the theory is defined, a {\em universal horizon} can be defined as a one-way membrane of
particles with arbitrarily large speeds, that is, a surface that traps all particles: once they get in, they shall be trapped inside the horizon all the time, even they have infinitely large speeds.
In the decoupling limit, Blas and Sibiryakov showed that such a surface indeed exists \cite{BS11}. Since then, universal horizons have been intensively studied \cite{UHs}.
In particular, to study universal horizons in gravitational theories without the presence of the \ae ther field, one way is to promote the \ae ther field to play the role as that of a Killing vector \cite{LACW}.

With the existence of universal horizons in static and stationary spacetimes \cite{UHs}, another natural question is: do universal horizons have a thermal interpretation, similar to Killing horizons in GR?
In the framework of the Einstein-\ae ther theory, it was argued that universal horizons should possess such properties, and showed explicitly that the first law of black hole mechanics indeed exists for static and neutral universal horizons \cite{BBM12}.
The definition of the surface gravity at the universal horizon is replaced by,
\bqn
\lb{1.1}
\begin{split}
  &\kappa_{\text{peeling}}=\left.\frac{1}{2}\frac{d}{dr}\frac{dr}{dv}\right|_{UH}=\left.\frac{1}{2}\frac{d}{dr}\left(\frac{s^r}{s^v}\right)\right|_{UH} \\
  & = \kappa_{UH} \equiv \left. \frac{1}{2}\nabla_{u}\left(u\cdot\chi\right)\right|_{UH},
\end{split}
\eqn
which was first obtained by considering the peeling behavior of a null ray near universal horizons \cite{CLMV}, and later was shown that the above relation holds for all static universal horizons \cite{LGSW}, where $u_{\mu}$ is the four-velocity of the \ae ther field, and $s_{\mu}$ is a spacelike vector orthogonal to $u_{\mu}$ and tangent to the universal horizons.
The quantities, $r$ and $v$, denote the Eddington-Finkelstein coordinates, in terms of which the static spacetimes are described
by the metric \cite{LGSW},
 \bqn
 \lb{1.2}
 ds^2 = - F(r) dv^2 + 2f(r) dv dr + r^2d\Sigma_k^2,
 \eqn
 where $ k = 0, \pm 1$, and
 \bqn
\lb{1.3}
 d\Sigma^2_k =\left\{
  \begin{array}{cc}
    d\theta^2+\sin^2\theta d\varpi^2,  & $k = 1$, \\
    d\theta^2+d\varpi^2,               & $k = 0$, \\
    d\theta^2+\sinh^2\theta d\varpi^2, & $k = -1$. \\
  \end{array}
\right.
 \eqn
It is interesting to note that the universal horizons always exist inside the Killing horizons in the well-known Schwarzschild, Schwarzschild-de Sitter and Reissner-Nordstr\"om spacetimes \cite{LGSW}, which are also solutions of Ho\v{r}ava gravity \cite{Wang17}.

It must be noted that it is still an open question how to generalize such a first law to charged and/or rotating universal horizons \cite{DWW,DWWZ,DLWJ,PLa,LPb}.
In addition, in the case of a neutral universal horizon, it was showed, via the tunneling approach, that the universal horizon radiates as a blackbody at a fixed temperature, even if the scalar field equations also violate local Lorentz invariance \cite{BBM13}, where the temperature $T_{UH}$ satisfies the canonical relation $T_{UH} = \kappa_{UH} /(2\pi)$.
However, if the Lorentz invariance is also allowed to be broken in the matter sector, such as in the case of Ho\v{r}ava gravity \cite{Horava09,Wang17}, the dispersion relation of a massive particle
will contain generically high-order momentum terms,
\bqn
\lb{1.4}
E^{2} = m^{2} +  c_{p}^2p^{2}\left(1 + \sum^{2(z-1)}_{n=1}{a_{n}\left(\frac{p}{M_{*}}\right)^{n}}\right),
\eqn
where $E$ and $p$ are the energy and momentum of the particle considered, and $c_p$ and $a_n$'s are coefficients, depending on the species of the particle, and $z$ is the dynamical exponent and denotes the power of the leading terms in the ultraviolet (UV) regime of the dispersion relation.
In Ho\v{r}ava gravity, the renormalizability of the theory requires $z \ge d$, where $d$ is the spatial dimension of the spacetime,
and $M_{*}$ is the suppression energy  scale of the higher-dimensional operators.
Then, it was shown that the universal horizon radiates still as a blackbody at a fixed temperature, but remarkably now it depends on $z$ \cite{DWWZ},
\bqn
\lb{1.5}
T_{UH}^{z\ge 2} = \frac{2(z-1)}{z}\left(\frac{\kappa_{UH}}{2\pi}\right),
\eqn
where $\kappa_{UH}$ is the surface gravity defined by Eq.(\ref{1.1}).
When  $z = 2$ we obtain the results of \cite{BBM13}, which considered only the case $z = 2$.
Recently, more careful studies of ray trajectories showed that the surface gravity for particles with a non-relativistic dispersion relation  (\ref{1.4}) is given by \cite{DL16},
\bqn
\lb{3.13}
\kappa_{UH}^{z\ge 2} = \left(\frac{2(z-1)}{z}\right) \kappa_{UH}.
\eqn
Similar results (with a factor 2 difference) were also obtained in \cite{Cropp16}.

In the study of black hole thermodynamics, specifically the first law, the energy of the system is usually identified as the ADM mass, and the entropy term  is identified as the (quasi-)local energy.
This was done initially by Brown and York, who applied the quasi-local energy to the black hole thermodynamics \cite{BY93}, whereby the first law was obtained.
Soon, Wald formulated it by using the Noether theorem \cite{wald94}.
Later, a new formula of quasi-local energy was proposed, and then applied to the study of black hole thermodynamics \cite{CNT95}.
From the above investigations, it can be seen that the quasi-local energy plays a key role in the understanding of the black hole thermodynamics \cite{Szabados}.

Applying the above conceptions to black holes in Einstein-\ae ther theory, Eling first considered, respectively, the Einstein and Landau-Lifshitz pseudo-tensors \cite{Eling},
while Foster investigated the same problem by using the Wald formalism \cite{Foster}.
From their expressions, one can deduce the ADM mass in asymptotically flat spacetimes with a modification due to the presence of the \ae ther field.
Recently, along the same line, Pacilio and Liberati obtained an integral form of the Smarr formula by assuming a scaling symmetry in  Einstein-\ae ther theory \cite{PLa}.
They also showed that the cosmological constant term can be also included, although it breaks the scaling symmetry.

In this letter, using the Wald formalism, we investigate the thermodynamics of charged black holes in D-dimensional stationary
spacetimes with or without rotations in Einstein-\ae ther-Maxwell theory. In particular, assuming the existence of a scaling symmetry of the 
action  \cite{PLa}, we obtain the Smarr integral formula, which can be applied to both Killing and universal horizons. When 
restricted to 4-dimensional spherically symmetric spacetimes, previous results obtained by a different method in \cite{DWW} are
re-derived. Specifically, in Sec. 2, we give a brief review over the Einstein-\ae ther-Maxwell theory, while in Sec. 3, applying Wald's formalism, 
we derive the integral form of the Smarr formulas for the Einstein-\ae ther-Maxwell theory in D-dimensional spacetimes with or without rotations.
In Sec. 4 we present our main conclusions and some general remarks and discussions.

\section{Einstein-\ae ther--Maxwell theory}
\renewcommand{\theequation}{2.\arabic{equation}} \setcounter{equation}{0}

The Einstein-\ae ther theory minimally coupled with an electromagnetic field is described by the action \cite{DWW},
\bqn \label{Lag}
\mathcal{S}=
\int d^4x\sqrt{-g}\Bigg[\frac{1}{16\pi G_{\aet}}\Big(\mathcal{R}-2\Lambda+\mathcal{L}_{\aet} +\mathcal{L}_M\Big)\Bigg]\,, \label{action}
\eqn
where $\mathcal{R}$ is the Ricci scalar, and $\Lambda$ is the  cosmological constant.
The   \ae ther Lagrangian $\mathcal{L}_{\aet}$ is given by
\bqn
\lag_{\aet}=-Z^{ab}_{~~cd}(\nabla_au^c)(\nabla_bu^d) +\lambda(g^{ab}u_au_b+1),
\eqn
where
the tensor $Z^{ab}_{~~cd}$ is defined as \cite{JM04}
\bqn \lb{tensor:Z}
Z^{ab}_{~~cd}=c_1g^{ab}g_{cd}+c_2\delta^a_{~c}\delta^b_{~d}
+c_3\delta^a_{~d}\delta^b_{~c}-c_4u^au^bg_{cd}\,,
\eqn
where $c_i (i = 1, 2, 3, 4)$ are coupling constants of the theory.
The   Newtonian constant $G_N$ is related to  $G_{\aet}$ via the relation $G_{\aet} =(1-c_{14}/2)G_N$ \cite{CL,Foster,Eling}.
There are a number of theoretical and observational bounds on the coupling constants $c_i$ \cite{JM01,EMS,yagi}. In particular,
 the combination of the gravitational wave event GW170817 \cite{GW170817}, observed by the LIGO/Virgo collaboration, and the one of the gamma-ray burst GRB 170817A \cite{GRB170817},
provides  much more severe constraint on $c_{13}$. In fact,  these events imply that the speed of the spin-2 mode $c_T$ must satisfy the bound, $- 3\times 10^{-15} < c_T -1 < 7\times 10^{-16}$. In the Einstein-aether theory, the speed of the spin-2 graviton is given by $c_{T}^2 = 1/(1-c_{13})$ \cite{JM04}, so the  GW170817 and GRB 170817A events imply
$\left |c_{13}\right| < 10^{-15}$. For the constraints on other parameters after the above observations are taken into account, see \cite{OMW18} for more details.
As to be shown below,   the formulas  developed in this letter do not impose any specific conditions on the choice of these free parameters,  so in principle they are valid with respect to any constraint.

Another useful expression of $\mathcal{L}_{\aet}$ is related to the irreducible decompositions of the  \ae ther field \cite{jacobson14},
\bqn
\nabla_au_b=\frac{1}{3}\theta h_{ab}+\sigma_{ab}+\omega_{ab}-u_aa_b,
\eqn
where $h_{ab}\equiv g_{ab}+u_au_b$, and
\bqn
\begin{split}
& \theta\equiv\nabla\cdot u, \quad a^a \equiv \nabla_uu^a \\
& \sigma_{ab}\equiv\nabla_{(a}u_{b)}+u_{(a}a_{b)}-\frac{1}{3}\theta h_{ab}, \\
&\omega_{ab}\equiv\nabla_{[a}u_{b]}+u_{[a}a_{b]},
\end{split}
\eqn
where $\nabla_X\equiv X^b\nabla_b$.
Note that, if the \ae ther satisfies the hypersurface-orthogonal condition, $u_{[a}\nabla_bu_{c]}=0$, then the twist $\omega_{ab}$ vanishes identically.
Then, we find that
\bqn
\begin{split}
{\cal{L}}_{u} &\equiv -Z^{ab}_{~~cd}(\nabla_au^c)(\nabla_bu^d) \\
              &=\frac{1}{3}c_{\theta}\theta^2+c_\sigma\sigma^2+c_\omega\omega^2-c_aa^2,
\end{split}
\eqn
where
\bqn  \lb{irr_ae}
\begin{split}
c_\theta &\equiv c_1+c_3+3c_2, \quad c_\sigma\equiv c_1+c_3, \\
c_\omega &\equiv c_1-c_3, \qquad c_a\equiv c_1+c_4.
\end{split}
\eqn
 The source-free Maxwell Lagrangian $\mathcal{L}_M$ is given by
\bqn
\mathcal{L}_M=-\mathcal{F}_{ab}\mathcal{F}^{ab}, \quad
~\mathcal{F}_{ab}=\nabla_a\mathcal{A}_b-\nabla_b\mathcal{A}_a,
\eqn
where $\mathcal{A}_a$ is the  four-vector  potential of the electromagnetic field.

To proceed  further, let us introduce the tetrad,  $\left(u^a, s^a, m^a, n^a\right)$, where  the space-like unit $m^a$ and $n^a$ lie on the tangent plane of the two-spheres $\mathcal{S}$, and $s^a$ is another space-like vector, which is orthogonal to
$u^a, \;  m^a$ and $n^a$. In terms of these four vectors,  the metric takes the form,
\bqn
 g_{ab} = -u_au_b + s_as_b + \hat{g}_{ab},
\eqn
where $\hat{g}_{ab}\equiv m^am^b+n^an^b$ is  the metric on the unit two-sphere $\mathcal{S}$. Due to the spherical symmetry, any vector $V^a$ can be decomposed as,
\bqn \label{vec}
 V^a=-(V\cdot u)u^a+(V\cdot s)s^a.
\eqn
In particular, the acceleration  has only one component along $s^a$ and is given by  $a^a=(a\cdot s)s^a$.
Additionally, any rank-two tensor $F_{ab}$ can have components along the directions of the bi-vectors $u_au_b,~u_{(a}s_{b)},~u_{[a}s_{b]},~s_as_b$ and $\hat{g}_{ab}$.
Especially, in \cite{BBM12, CroppLV} the following useful relations were given,
\bqn
\label{def:CoVa}
\nabla_a u_b &=& -(a\cdot s)s_b u_a + K_0s_b s_a +
\frac{1}{2}\hat{K}^{u}\hat{g}_{ab}, \\
\label{def:CoVb}
\nabla_a s_b &=& -(a\cdot s)u_b u_a + K_0 u_b s_a  +
\frac{1}{2}\hat{K}^s\hat{g}_{ab},
\eqn
where $\hat{K}^{u}$ and $\hat{K}^{s}$ are given by,
\bqn
\label{KuKs}
\hat{K}^{u} &=& [m^c\nabla_c+n^c\nabla_c]^u, \nb\\
\hat{K}^{s} &=& [m^c\nabla_c+n^c\nabla_c]^s,
\eqn
where superscript $u$ and $s$ represent that the extrinsic curvature operator act on the vector $u_a$ and $s_a$ respectively.

The  Maxwell field $\mathcal{F}^{ab}$ is usually  decomposed into
  the electric   and magnetic  fields $E^a$ and $B^a$, respectively,
\bqn
E^a=\mathcal{F}^{ab}u_b, \quad B^a=\frac{e^{abmn}}{2\sqrt{-g}}\mathcal{F}_{mn}u_b,
\eqn
where $e^{abmn}$ is the Levi-Civita tensor. However, because of the spherical symmetry, we have   $B^a=0$, and then we find
\bqn
\mathcal{F}^{ab}=-E^au^b+E^bu^a.
\eqn
Because the electric field is spacelike, we have $(E\cdot u)=0$ and  $E^a=(E\cdot s)s^a$.
Thus, $\mathcal{F}_{ab}=-(E\cdot s)\hat\epsilon_{ab}$, where $\hat\epsilon_{ab}=(-u_as_b+u_bs_a)$.
 Then, from the vacuum Maxwell equations, we find $(E\cdot s)=Q/r^2$, where $r$ denotes the geometric radius of the two-sphere, and
 $Q$ is an integration constant, which represents the total charge of the whole spacetime. Therefore, we have
\bqn
\label{Maxwellb}
\mathcal{F}_{ab}=-\frac{Q}{r^2}\hat\epsilon_{ab}.
\eqn

The equations of motion, obtained by varying  the action (\ref{action}), respectively,  with respect to $g_{ab}$, $u^a$, $\mathcal{A}_a$ and $\lambda$ are given by
\bqn \label{motion}
\begin{split}
 & \mathcal{G}_{ab}+\Lambda
 g_{ab}=\mathcal{T}^{\aet}_{ab}+8\pi G_{\aet}\mathcal{T}^M_{ab}, \\
 & {\AEt}_a=0,   \\
 & \nabla_a\mathcal{F}^{ab}=0, \\
 & g^{ab}u_au_b=-1,
\end{split}
\eqn
where $\mathcal{G}_{ab}$ is the Einstein tensor, the \ae ther and Maxwell energy-momentum  tensors $\mathcal{T}^{\aet}_{ab}$ and $\mathcal{T}^M_{ab}$ are given by
\bqn  \label{EMTs}
\mathcal{T}^{\aet}_{ab}&=&c_1[(\nabla_au_c)(\nabla_bu^c)
-(\nabla_cu_a)(\nabla^cu_b)]+c_4a_aa_b  \nb\\
\quad &&+\nabla_cX^c_{~~ab}+\lambda u_au_b-\frac{1}{2}g_{ab}Y^c_{~~d}\nabla_cu^d
, \\
\mathcal{T}_{ab}^M
 &=&\frac{1}{16\pi G_{\aet}}\Big[-\frac{1}{4}g_{ab}\mathcal{F}_{mn}\mathcal{F}^{mn}
 +\mathcal{F}_{am}\mathcal{F}_{b}^{~m}\Big],
\eqn
with
\bqn \lb{tensor:YX}
{\AEt}_a&=&\nabla_bY^b_{~~a}+\lambda u_a+c_4(\nabla_au^b)a_b,  \\
 X^c_{~~ab}&=&Y^c_{~~(a}u_{b)}-u_{(a}Y^{~~c}_{b)}+u^cY_{(ab)}, \\
 Y^a_{~~b}&=&Z^{ac}_{~~~bd}\nabla_cu^d.
\eqn

\section{Smarr Formulas for Einstein-\ae ther--Maxwell theory}
\renewcommand{\theequation}{3.\arabic{equation}} \setcounter{equation}{0}

In this section, to have our results  be as much applicable as possible, we shall consider D-dimensional spacetimes in Einstein-\ae ther--Maxwell theory.
The only assumption is the existence of a Killing vector $\xi$ that describes the scaling symmetry.

For any  diffeomorphism-invariant $D$-form Lagrangian, one can always associate a local symmetry with its corresponding conservation quantities, i.e. a Noether current $(D-1)$-form {\bf J} as well as a Noether charge $(D-2)$-form {\bf Q}
\cite{wald94,Foster}. If such a symmetry is represented by a   Killing vector field, $\xi$, then the corresponding  Noether current $(D-1)$-form {\bf J} is given by
\bqn \label{J1}
\J[\xi]=\bf\Theta(\varphi,\Lie_\xi\varphi)-\xi\cdot\Lag,
\eqn
where $\Lie_\xi$ is the Lie derivative along the $\xi$ direction, and {\bf J} is closed on shell \footnote{Here, ``$\simeq$'' represent the equality when the field equation, $\textbf{E}=0$, hold on shell.},
\bqn
\ud\J[\xi]=-\textbf{E}\Lie_\xi\varphi\simeq 0,
\eqn
which implies  that corresponding to $\xi$ there is a $(D-2)$-form Noether charge, $\Q$,
\bqn \label{J2}
\J[\xi]\simeq\ud\Q[\xi].
\eqn
 In the Einstein-\ae ther--Maxwell theory, the Noether charge is given by
\bqn
\label{eq:ae:Q}
\begin{split}
\Q[\xi]=\frac{-1}{16\pi G_{\aet}}&\left[\nabla^a\xi^b+u^a Y^b{}_c\xi^c+u^a Y_c{}^b\xi^c \right.  \\
       &\qquad \left.+Y^{ab}\ux+2\mathcal{F}^{ab}\mathcal{A}_c\xi^c\right]\epsilon_{ab}.
\end{split}
\eqn
For a solution $\{\varphi\}$ of the theory, when it is invariant along $\xi$, that is, $\Lie_\xi\varphi\simeq 0$,  we have  ${\bf\Theta}(\varphi,\Lie_\xi\varphi)\simeq 0$. From (\ref{J1}) and (\ref{J2}), we integrate the expression,
$\J[\xi]+\xi\cdot\Lag\simeq 0$, over an hypersurface $\Sigma$ and its boundary, $\partial\Sigma$, respectively, we find that
\bqn
\lb{OnShell}
\int_{\partial\Sigma}\Q[\xi]+\int_\Sigma\xi\cdot\Lag = 0.
\eqn
In \cite{LP}, it was observed that Eq.(\ref{OnShell}) leads to the Smarr formula for sufficiently general diffeoinvariant theories, while in \cite{PLa} it was shown that scaling symmetry leads the Lagrangian of Einstein-Aether theory
to be a total divergence on-shell,   i.e. $\Lag\simeq\ud\B$, with a $(D-1)$-form $\B$ \footnote{In \cite{PLa}  the Smarr formulas for the infrared Horava gravity were also studied, and found that they reduce to a surface integral only when the integrals are evaluated in the preferred frame.}.
Then we can derive the Smarr formula by taking a surface integral under a flow of the Killing vector field $\xi$. For the action (\ref{Lag}), the scaling symmetry is 
\bqn \label{ST}
\begin{split}
  &g_{ab}\rightarrow\gamma g_{ab} \\
  &u_a\rightarrow\gamma^{-\frac{1}{2}} u_a \\
  &\lambda\rightarrow \gamma^{-1}\lambda \\
  &\mathcal{A}^a\rightarrow \gamma^{\frac{1}{2}}\mathcal{A}^a,
\end{split}
\eqn
where $\gamma$ is a constant. Following   \cite{PLa,LP}, we find
\bqn
\label{eq:ae:Lshell}
\begin{split}
&\mathcal{R}+\lag_u+\lag_M  \\
&=\left(\frac{2}{D-2}\right)\nabla_m(u^aY_a{}^m-u^mY^a{}_a-2\mathcal{F}^{ma}\mathcal{A}_a).
\end{split}
\eqn
Then,  total Lagrangian,
$\Lag=\mathcal{R}+\lag_\aet+\lag_M$, becomes a total divergence on shell, i.e. $\Lag\simeq\ud\B$, where
\bqn
\label{eq:ae:B}
\B=\left(\frac{2}{D-2}\right)\frac{1}{16\pi G_{\aet}}\left(u^aY_a{}^m-u^mY^a{}_a-2\mathcal{F}^{ma}\mathcal{A}_a\right)\epsilon_m.
\eqn
Thus, on shell, Eq.(\ref{OnShell}) reduces to a surface integral,
\bqn
\label{eq:smarr:B}
0\simeq\int_{\partial\Sigma}\Q[\xi]-\xi\cdot\B.
\eqn

\subsection{The Expression for $\Q[\xi]-\xi\cdot\B$}

Inserting the expressions of $Z^{ab}{}_{cd}$ given by (\ref{tensor:Z}) and $Y^a{}_b$, $X^a{}_{bc}$ given by (\ref{tensor:YX}) into  (\ref{eq:ae:Q}) and (\ref{eq:ae:B}), we find
\bqn
 \lb{eq:QB1}
\begin{split}
 \textbf{Q}[\xi]-&\xi\cdot \B = -\frac{1}{16\pi G_{\aet}}\Bigg[\nabla^a\xi^b -2c_4u^aa^b(u\cdot\xi)\\
 & + 2c_{13}u^a\xi_c\nabla^{(b}u^{c)}+(c_1-c_3)\nabla^au^b(u\cdot\xi) \\
 & -\frac{2}{D-2}(c_{14}\xi^aa^b+c_{123}u^a\xi^b(\nabla\cdot u)) \\
 & +2\left(F^{ab}\A_c\xi^c-\frac{2}{D-2}\F^{ac}\A_c\xi^b\right)\Bigg]\epsilon_{ab}.
\end{split}
\eqn
Alternatively, in terms of the fluid coefficients of Eq.(\ref{irr_ae}), we find
\bqn
  \label{eq:NQ1}
\begin{split}
\textbf{Q}[\xi]-& \xi\cdot \textbf{B} = -\frac{1}{16\pi G_{\aet}}\Bigg[\nabla^a\xi^b -2c_a(u\cdot\xi)u^aa^b\\
& + c_{\omega}(u\cdot\xi)\omega^{ab}  +2c_{\sigma}u^a\xi_c\nabla^{(b}u^{c)} + c_{\sigma}(u\cdot\xi)u^aa^b \\
& -\frac{2}{D-2}\left(c_{123}u^a\xi^b(\nabla\cdot u) + c_a\xi^aa^b \right) \\
& +2\left(\F^{ab}\A_c\xi^c-\frac{2}{D-2}\F^{ac}\A_c\xi^b\right)\Bigg]\epsilon_{ab}.
\end{split}
\eqn

It should be noted that the above expressions hold for any stationary spacetimes, including the ones with rotations.

\subsection{4-Dimensional Spherically Symmetric Spacetimes}

In \cite{DWW}, it was shown that, using  the Einstein-aether-Maxwell  field equations,  the geometric identity \cite{BCH},
\begin{eqnarray}
\label{ID}
\mathcal{R}_{ab}\xi^b=\nabla^b(\nabla_a\xi_b),
\end{eqnarray}
can be written in the  form,
\begin{eqnarray}
\label{SmarrFa}
\nabla_bF^{ab}=0,\;\;\; F^{ab} \equiv 2F(r)u^{[a}s^{b]},
\end{eqnarray}
where $F(r) =  F^Q(r) + q(r)$, and
\begin{eqnarray} \label{SmarrFb}
\begin{split}
& \nabla^b\left(F^Q(r) u_{[a}s_{b]}\right) =  - \frac{Q^2}{2r^4}\xi_a, \\
& q(r)  \equiv    -\left(1-\frac{c_{14}}{2}\right)(a\cdot s)(u\cdot \xi) \\
& ~~~~~~~~~~~~~ +\left[(1-c_{13})K_0 +\frac{c_{123}}{2}K\right](s\cdot \xi).
\end{split}
\end{eqnarray}

Then, using Gauss' law,  from Eq.(\ref{SmarrFa}) we find that
\begin{eqnarray}\label{tmassB}
   0 &=& \int_{\Sigma}{\left(\nabla_bF^{ab}\right)   d\Sigma_a} =   \int_{{\cal{B}}_{\infty}}{F^{ab}   d\Sigma_{ab}} - \int_{{\cal{B}}_{H}}{F^{ab}   d\Sigma_{ab}}\nb\\
   &=&   \int_{{\cal{B}}_{\infty}}{FdA} - \int_{{\cal{B}}_{H}}{F dA}.
\end{eqnarray}
Here $d\Sigma_a$ is the surface element of a spacelike  hypersurface $\Sigma$. The boundary $\partial\Sigma$ of $\Sigma$
consists of the boundary  at spatial infinity ${\cal{B}}_{\infty}$,  and the horizon ${\cal{B}}_{H}$, either the Killing or the universal.

In the following, we shall show that Eq.(\ref{tmassB}) can be obtained from Eq.(\ref{eq:smarr:B}), when it is restricted to 4-dimensional spherically symmetric spacetimes.
To this goal, let us first note that in the sperically symmetric spacetimes the hypersurface-orthogonal condition  is automatically satisfied, so the twist $\omega^{ab}$ vanishes identically,
while $\epsilon_{ab}$ can be written as   $\epsilon_{ab}=\hat\epsilon_{ab}\bar\epsilon$. Inserting Eq.(\ref{Maxwellb}) into
Eq. (\ref{eq:NQ1}), we find
\bqn
\begin{split}
 \textbf{Q}[\xi] -& \xi\cdot \textbf{B} = -\frac{1}{16\pi G_{\aet}}\Bigg[\nabla^a\xi^b\hat\epsilon_{ab}
+ 2c_{\sigma}(s\cdot\xi)s^as^b\nabla_au_b \\
&-\frac{2}{D-2}\left[c_{123}(s\cdot\xi)(\nabla\cdot u) - c_a(u\cdot\xi)(a\cdot s)\right] \\
& -2c_a(u\cdot\xi)(a\cdot s)+ c_{\omega}(u\cdot\xi)\omega^{ab}\hat{\epsilon}_{ab} \\
& +\frac{2Q}{r^2}\left(-(\xi\cdot\A)\hat{\epsilon}^{ab}
  +\frac{2}{D-2}\A_c\xi^b\hat{\epsilon}^{ac}\right)\hat{\epsilon}_{ab}\Bigg]\bar\epsilon, \nb
\end{split}\\
\eqn
where $\xi^a=-(u\cdot\xi)u^a+(s\cdot\xi)s^a$ and $\A$ will be defined below.
Finally, for $D=4$, with null twist condition and $(\nabla^bu^a+u^ba^a)\hat{\epsilon}_{ab}=0$,  we obtain
\bqn \label{eq:QB2}
\begin{split}
\Q[\xi] &- \xi\cdot \B =
\frac{1}{8\pi G_{\aet}}\Bigg[-\left(1-\frac{c_a}{2}\right)(u\cdot\xi)(a\cdot s) \\
& + (1-c_{\sigma})(s\cdot\xi)s^as^b\nabla_au_b +\frac{c_{123}}{2}(s\cdot\xi)(\nabla\cdot u) \\
& -\frac{Q}{r^2}[2(\xi\cdot\A)+(u\cdot\A)(u\cdot\xi)-(s\cdot\A)(s\cdot\xi)]\Bigg]\bar\epsilon,
\end{split} \nb\\
\eqn
where we had used (\ref{def:CoVa}) to get the first term,
\bqn
\nabla^a\xi^b\hat{\epsilon}_{ab}=2[(u\cdot\xi)(a\cdot s)-(s\cdot\xi)K_0],
\eqn
 in r.h.s. of Eq.(\ref{eq:QB2}).
With the above expression for $\Q[\xi]-\xi\cdot \B$ and  by taking the integral of Eq.(\ref{eq:smarr:B})  over the  boundary  $\partial\Sigma=\Sigma_\infty\cup \Sigma_{\text{BH}}$,
we find
\bqn
\lb{SFs}
\int_{\Sigma_\infty}\left(\Q[\xi]-\xi\cdot\B\right)- \int_{\Sigma_{\text{BH}}}\left(\Q[\xi]-\xi\cdot\B\right) \simeq 0,
\eqn
which is nothing but Eq.(\ref{tmassB}).

\section{Conclusion}

In this work, applying  the Wald formalism to the Einstein--\ae ther--Maxwell theory, we have derived the Smarr integral formula in D-dimensional stationary spacetimes with or without
rotations, by assuming the existence of  a scaling symmetry. Applying it to Killing or universal horizons, one can obtain the integral form of the first-law of thermodynamics
 in Einstein--\ae ther--Maxwell theory.  Restricting to 4-dimensional spherically symmetric spacetimes, we have  re-derived   the Smarr integral formulae  for charged black holes obtained in \cite{DWW}.


This work was supported in part by National Natural Science Foundation of China (NNSFC) with the grant  Nos.: 11375153 (A.W.), 11675145 (A.W.),  11475148 (H.S.L.), 11675144 (H.S.L.), and 11605155 (S.J.Z).

\end{document}